\documentclass[twocolumn]{aastex62}
\usepackage{ulem}
\usepackage{tablefootnote}

\graphicspath{{./}{figures/}}
\def\Msun{M_\odot}

\begin{document}

\title{The Effect of Warm Dark Matter on Early Star Formation Histories of Massive Galaxies: Predictions from the CROC Simulations}

\correspondingauthor{Clarke Esmerian}
\email{cesmerian@uchicago.edu}

\author{Clarke J.\ Esmerian}
\affiliation{Department of Astronomy \& Astrophysics, University of Chicago\\
Chicago, IL 60637 USA}
\affiliation{Kavli Institute for Cosmological Physics;
The University of Chicago;
Chicago, IL 60637 USA}
\affiliation{Fermi National Accelerator Laboratory;
Batavia, IL 60510, USA}

\author{Nickolay Y.\ Gnedin}
\affiliation{Fermi National Accelerator Laboratory;
Batavia, IL 60510, USA}
\affiliation{Kavli Institute for Cosmological Physics;
The University of Chicago;
Chicago, IL 60637 USA}
\affiliation{Department of Astronomy \& Astrophysics, University of Chicago\\
Chicago, IL 60637 USA}

\begin{abstract}
Several massive ($M_{*} > 10^8 M_{\odot}$), high-redshift ($z = 8-10$) galaxies have recently been discovered to contain stars with ages of several hundred million years, pushing the onset of star formation in these galaxies back to $z\sim15$. The very existence of stars formed so early may serve as a test for cosmological models with suppressed small-scale power (and, hence, late formation of cosmic structure). We explore the ages of the oldest stars in numerical simulations from the Cosmic Reionization On Computers (CROC) project with cold dark matter (CDM) and two warm dark matter (WDM) cosmologies with 3 and 6 keV particles. There are statistically significant differences of $\sim 5\;{\rm Myr}$ between average stellar ages of massive galaxies in CDM and 3 keV WDM, while CDM and 6 keV WDM are statistically indistinguishable. Even this 5 Myr difference, however, is much less than current observational uncertainties on the stellar population properties of high-redshift galaxies. The age distributions of all galaxies in all cosmologies fail to produce a substantial Balmer break, although uncertainties in dust attenuation are a potentially significant factor. Finally, we assess the convergence of our simulation predictions and find that the systematic uncertainties on individual galaxy properties are comparable to the differences between cosmologies, suggesting these differences may not be numerically robust.
\end{abstract}




\section{Introduction}

Indirect observational evidence for dark matter is incontrovertible. The standard cosmological model has been extremely successful in explaining observational data on the cosmic microwave background \citep[e.g.][]{Planck2018VI}, large-scale structure \citep[e.g.][]{SDSSDR12Cosmo}, and galaxy formation \citep[see, e.g.][]{NaabOstrikerARAA2017}, invariably requiring that $\sim 25\%$ of the universal mass-energy density behaves like a collisionless, nonrelativistic, electrically neutral, and massive particle, at least on scales $\gtrsim$ kpc. Decades of cosmological, astrophysical, and direct-detection probes have failed to present compelling evidence for any modification to this picture.

During the cosmic expansion dark matter particles diffused on the free-streaming scale defined by their characteristic velocity, smoothing density perturbations and consequently suppressing power on small scales. The resulting cutoff in the power spectrum of density fluctuations can be directly related to the minimum mass that collapsed to form gravitationally bound dark matter halos. A broad class of physically well-motivated, viable candidates termed warm dark matter (WDM) decoupled with sufficiently large velocities that this minimum dark matter halo mass was comparable to the masses of the first galaxies. The average dark matter particle velocity is inversely related to the dark matter particle mass $m_x$, implying a direct relation between this mass and the scale of suppressed power in these models.

Current constraints on the small-scale dark matter power spectrum, translated to a 2$\sigma$ lower bound on the WDM particle mass, give $m_x \gtrsim 2.4\, {\rm keV}$ from the high-$z$ galaxy luminosity function \citep{Menci2016}, $m_x \gtrsim 3\, {\rm keV}$ from the Ly$\alpha$ forest \citep{Irsivic2017, Yeche2017}, and $m_x \gtrsim 5.2\, {\rm keV}$ from strong gravitational lensing \citep{Gilman2019}. Future observations of the stellar mass-halo mass relation \citep{Kang2013}, halo bias \citep{Carucci2015}, and Milky Way satellite galaxy population \citep{Kennedy2014, Kang2020} may provide additional constraints. However, each of these methods is subject to substantial astrophysical systematic uncertainties and a thermal relic mass of $ m_x \gtrsim 5\, {\rm keV}$ remains consistent with all observational constraints, motivating a careful look at unexplored astrophysical probes that might further constrain models in this parameter space.

The star formation histories of the first galaxies are another such probe. This is because they are ultimately determined by the gravitational collapse of the first dark matter halos. Recent observational analyses \citep{Hashimoto2018, Strait2020} have claimed evidence for the onset of star formation within the first few hundred million years after the Big Bang, indicating that we may be starting to detect the onset of galaxy formation. Furthermore, the data from this cosmic epoch obtained by the James Webb Space Telescope will facilitate the statistical characterization of early galaxy populations. Motivated by these current and anticipated future observational constraints, in this paper we assess the extent to which high-redshift massive galaxy star formation histories are sensitive to the properties of dark matter.


\begin{figure*}
    \includegraphics[width=\textwidth]{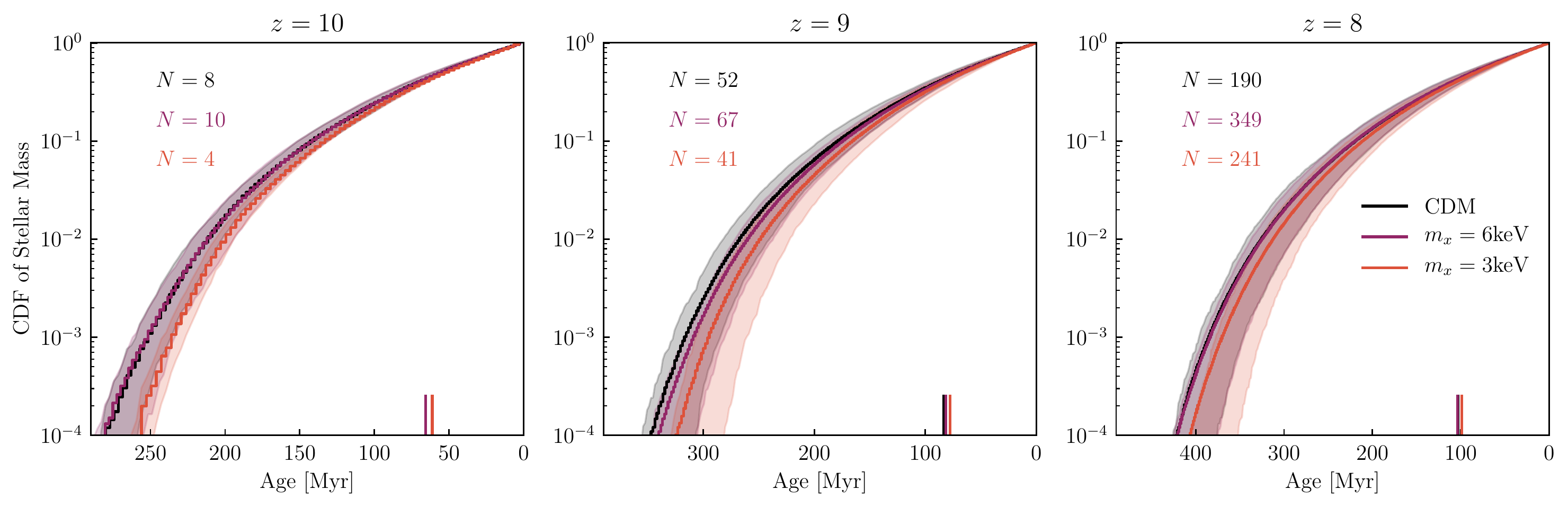}
    \caption{Comparison of stellar age distributions. CDFs of star particle ages for the most massive galaxies ($M_* > 10^8 M_{\odot}$) at high redshifts in CROC simulations with different dark matter cosmologies. The number of galaxies above this threshold is shown in the upper left corner legend of each top panel. Colors label the three dark matter models. Solid lines indicate the stellar age CDF of all galaxies above the stellar mass threshold combined. Shaded regions show the 16th and 84th percentiles at each age for individual galaxy CDFs. Cosmologies with a power spectrum cutoff predicted by a $3 {\rm keV}$ thermal relic display delayed star formation compared to CDM.}
    \label{fig:age_CDFs}
\end{figure*}

\section{Simulations: The Cosmic Reionization on Computers Project}

\subsection{General Description}
As a primary tool for our investigations we use simulations from the Cosmic Reionization on Computers (CROC) project \citep{CROCI, CROCII}. These simulations use the cosmological $N$-body + adaptive mesh gas dynamics code ART \citep{Kravtsov1999PhDT, Kravtsov2002, Rudd2008} to
account for cosmology, gravity, gas dynamics coupled to radiative transfer, radiative cooling, star formation, and stellar feedback to model the Epoch of Reionization.
We analyze simulations from the ``Caiman" series presented in \citet{Gnedin2017}, which are an update of the initial CROC methodology calibrated to demonstrate weak numerical convergence as determined in \citet{CROC_Convergence}. Details of the numerical methods and physical models employed in these simulations can be found in the above references.

\subsection{Warm Dark Matter}
\label{sec:calibration}
WDM effects on reionization with the CROC simulations were first explored in \citet{CROC_WDM}, wherein the galaxy UV luminosity function, IGM opacity, and power spectra of redshifted 21 cm emission were compared between simulations with cold dark matter (CDM) and 3 keV WDM cosmologies. We extend this work by investigating star formation histories in a larger suite of simulations. WDM is accounted for in the same way as in \citet{CROC_WDM}, namely setting Gaussian-random initial conditions with a CDM power spectrum, filtered by a transfer function that suppresses power at high $k$:

\begin{equation}
T_{\rm WDM}(k) = [1 + (\alpha k)^{2\nu}]^{-5/\nu}
\end{equation}

\noindent where the parameters are taken to be
\begin{equation}
\nu = 1.12
\end{equation}
\begin{equation}
\alpha = 0.049\left(\frac{m_x}{1{\rm keV}}\right)^{1.11}\left(\frac{\Omega_x}{0.25}\right)^{0.11}\left(\frac{h}{0.7}\right)^{1.22} {\rm Mpc}\;h^{-1}
\end{equation}

\noindent based on fits obtained from numerical solutions to the Boltzmann equation presented in \citet{Viel2005}.

The challenge in comparing simulations of different cosmologies is in ensuring that the comparison is fair. Galaxy formation simulations include several numerical parameters that control the subgrid models of star formation and stellar feedback. These parameters are usually calibrated against a subset of observational data, since they cannot be predicted from first principles at computationally feasible resolutions. In the case of CROC simulations, star particles are created by assuming a constant depletion timescale $\tau_{\rm SF}$ of molecular gas, stellar feedback in the form of thermal energy from supernova explosions is implemented with a delayed cooling timescale $\tau_{\rm BW}$, and ionizing radiation from stellar populations is attenuated by an escape fraction factor $\epsilon_{\rm UV}$. 

For the CDM case these parameters ($\tau_{\rm SF}$, $\tau_{\rm BW}$, $\epsilon_{\rm UV}$) were calibrated during the design stage of the CROC project to match the observed high-redshift galaxy UV luminosity function and the cosmic neutral hydrogen fraction as a function of redshift. Using the same values of these parameters in a WDM cosmology would break the calibration and make the WDM simulation appear to be inconsistent with observational constraints, but such inconsistency cannot be interpreted as a failure of the model without proper recalibration.

Such recalibration for 3 keV WDM cosmology was done in \citet{CROC_WDM}, where the star formation efficiency was increased by a factor of 1.5 (by decreasing the depletion time of molecular gas by the same factor). Since such a recalibration for 6 keV WDM was computationally infeasible, we employ this same star formation efficiency in the 6 keV WDM simulations presented here. Hence, this simulation is slightly out of calibration, producing systematically higher high-redshift UV luminosity functions than either CDM or 3 keV WDM (see Appendix~\ref{sec:appendix_calibration}). However, this difference is far less than the uncertainties in the observations with which the simulations are calibrated, so this simulation should be sufficiently accurate for our analysis.

\subsection{Simulation Volumes}
In order to have a reasonable sample of rare galaxies similar to the observed ones, we use a special subset of CROC simulations that model a significantly overdense region. Algorithmically, this is achieved via a nonzero value of the mean overdensity in the simulation volume, commonly referred to as a ``DC mode.'' The DC mode enhancement, described in \citet{Gnedin2011_DC}, is quantified by the value of the linear density fluctuation on the box scale. In this work we use a $3\sigma$ fluctuation, about as large as would make sense in our setup (the most massive halo in such a volume would be a progenitor of a $\sim3\times10^{15}\Msun$ cluster). 

We label these simulations ``B40 E.DC=3,'' all in a volume with $40h^{-1}\approx 60$ comoving Mpc on a side. For our primary analysis, we use three such simulations. These constitute realizations of CDM, 3 keV WDM, and 6 keV WDM cosmologies described above, run with $1024^3$ dark matter particles with the effective total mass resolution of $7\times10^6\Msun$ (we call this resolution ``standard'', as it has been used for most of CROC simulations). 

Additionally, in Section~\ref{sec:cos_var} we compare the predictions of 16 more simulations: six independent realizations of CDM cosmology with $40h^{-1}$ comoving Mpc box sizes (B40 A-F), six more with $20h^{-1}$ comoving Mpc boxes (B20 A-F), and the four 3 keV WDM $20h^{-1}$ comoving Mpc boxes (B20 A, B, D, and E) presented in \citet{CROC_WDM}. All of these were run with the standard resolution. For numerical convergence studies (see Section~\ref{sec:convergence}), we also employ an additional simulation that models CDM cosmology with $2048^3$ particles at the effective total mass resolution of $9\times10^5\Msun$ (alias ``high resolution"). 

\begin{deluxetable*}{|c|ccc|ccc|ccc|}
\setlength{\tabcolsep}{1.5pt}
\tablecaption{Mass-weighted Mean and Standard Deviation (St.D.) Star Particle Ages (in Myr) for Dark Matter Halos with Total Stellar Mass $M_* > 10^8M_{\odot}$\label{table:ages}.}
\startdata
\tablehead{
    \textbf{Cosmology, Volume}
    & 
    \colhead{$\bf z = 10$} & 
    &
    &
    \colhead{$\bf z = 9$} & 
    & 
    &
    \colhead{$\bf z = 8$} & &
    }
\textbf{CDM} & Mean & St.D. &  & Mean & St.D. &  & Mean & St.D. & \\
\hline
{\bf B40, E.DC=3} & 65.7 & 4.9 & $(N=8)$ &  83.4 & 8.1 & $(N=52)$ &  103.1 & 10.6 & $(N=190)$ \\
B40, A & & & &  87.9 & 2.7 & $(N=3)$ &  108.5 & 10.8 & $(N=16)$  \\
B40, B & 65.7 & & $(N=1)$ &  86.9 & 2.5 & $(N=5)$ &  106.1 & 10.6 & $(N=49)$  \\
B40, C & & & &  88.6 & 8.9 & $(N=14)$ &  108.1 & 10.4 & $(N=81)$  \\
B40, D & & & &  76.6 & 4.6 & $(N=4)$ &  102.1 & 9.6 & $(N=42)$  \\
B40, E & & & &  83.9 & 6.2 & $(N=13)$ &  104.6 & 10.0 & $(N=57)$  \\
B40, F & & & &  83.7 & 8.6 & $(N=4)$ &  105.3 & 9.5 & $(N=20)$  \\
B20, A & & & &  & & &  104.9 & 7.8 & $(N=5)$ \\
B20, B & & & &  & & &  107.4 & 4.6 & $(N=3)$  \\
B20, C & & & &  85.1 & & $(N=1)$ &  106.9 & 5.1 & $(N=6)$  \\
B20, D & & & &  78.1 & & $(N=1)$ &  107.3 & 2.9 & $(N=2)$  \\
B20, E & & & &  89.4 & 5.7 & $(N=2)$ &  104.9 & 10.5 & $(N=14)$ \\
B20, F & & & &  & & &  105.6 & 4.7 & $(N=4)$  \\
\hline
\textbf{6 keV WDM} & & & & & & & & & \\
\hline
{\bf B40, E.DC=3} & 65.7 & 4.5 & $(N=10)$ &  81.6 & 7.1 & $(N=67)$ &  103.3 & 10.4 & $(N=349)$ \\
\hline
\textbf{3 keV WDM} & & & & & & & & & \\
\hline
{\bf B40, E.DC=3} & 61.2 & 2.7 & $(N=4)$ &  77.6 & 8.3 & $(N=41)$ &  99.0 & 10.4 & $(N=241)$  \\
B20, A & & & &  & & &  92.8 & 8.3 & $(N=9)$  \\
B20, B & & & &  & & &  97.2 & 5.2 & $(N=3)$  \\
B20, D & & & &  71.6 & & $(N=1)$ &  97.8 & 3.8 & $(N=2)$  \\
B20, E & & & &  79.4 & 4.7 & $(N=3)$ &  96.3 & 11.6 & $(N=16)$  \\
\enddata
\end{deluxetable*}

\section{Results}
We first present a comparison between the star formation histories of the most massive galaxies between different assumed dark matter cosmologies in the CROC simulations (Section~\ref{sec:ages}), and then a comparison of these histories to current observational constraints (Section~\ref{sec:obs}). In the following analysis we focus on the most massive galaxies ($M_* > 10^8 M_{\odot}$) in three high redshift bins ($z = 8, 9$, and $10$). To avoid double-counting stars in overlapping halos or subhalos, we assign star particles to their most massive parent halo. Stellar masses are calculated as the mass of star particles within the virial radius of the dark matter halo, as identified by the ROCKSTAR halo finder \citep{ROCKSTAR}, or in the case of the high-resolution simulation the HOP halo finder \citep{HOP}, since running ROCKSTAR on those simulations was computationally infeasible.  

\subsection{Distribution of Stellar Ages}
\label{sec:ages}
We first examine the distribution of star particle ages for massive galaxies at high redshift in the three simulated dark-matter cosmologies, shown in Figure \ref{fig:age_CDFs}. In all cases, the average galaxy ages appear to be older in CDM than in 3 keV WDM, while the distribution of stellar particle ages appear to be almost identical between CDM and 6 keV WDM. This difference between CDM and 3 keV WDM is unsurprising because the nonnegligible free-streaming scale in WDM implies a later time at which the first dark matter halos collapse. It is nonetheless a nontrivial result, because the simulations were calibrated to reproduce observations of the high-redshift galaxy luminosity function and IGM neutral fraction in both cosmologies, and thereby equally well reproduce observational constraints on the star formation history of massive galaxies at high redshift.

\subsubsection{Cosmic Variance}
\label{sec:cos_var}
Table \ref{table:ages} quotes the mass-weighted mean stellar age of massive galaxies in all available simulations to assess the effect of cosmic variance on our simulation predictions. The number of galaxies associated with each volume mean are quoted. While these extra volumes only add an appreciable number of additional massive galaxies by $z=8$, they display a trend consistent with the E.DC=3 volume. Namely, the stellar ages of massive galaxies in 3 keV WDM are $\sim 5$ Myr younger than those in CDM. This suggests that the conclusions drawn from the E.DC=3 simulation about the differences between different cosmologies are robust to cosmic variance, so we focus exclusively on this simulation volume for the remainder of the analysis.

\subsection{Comparison to observations}
\label{sec:obs}
\subsubsection{SED-fitting-derived Ages}

\begin{figure*}
    \includegraphics[width=\textwidth]{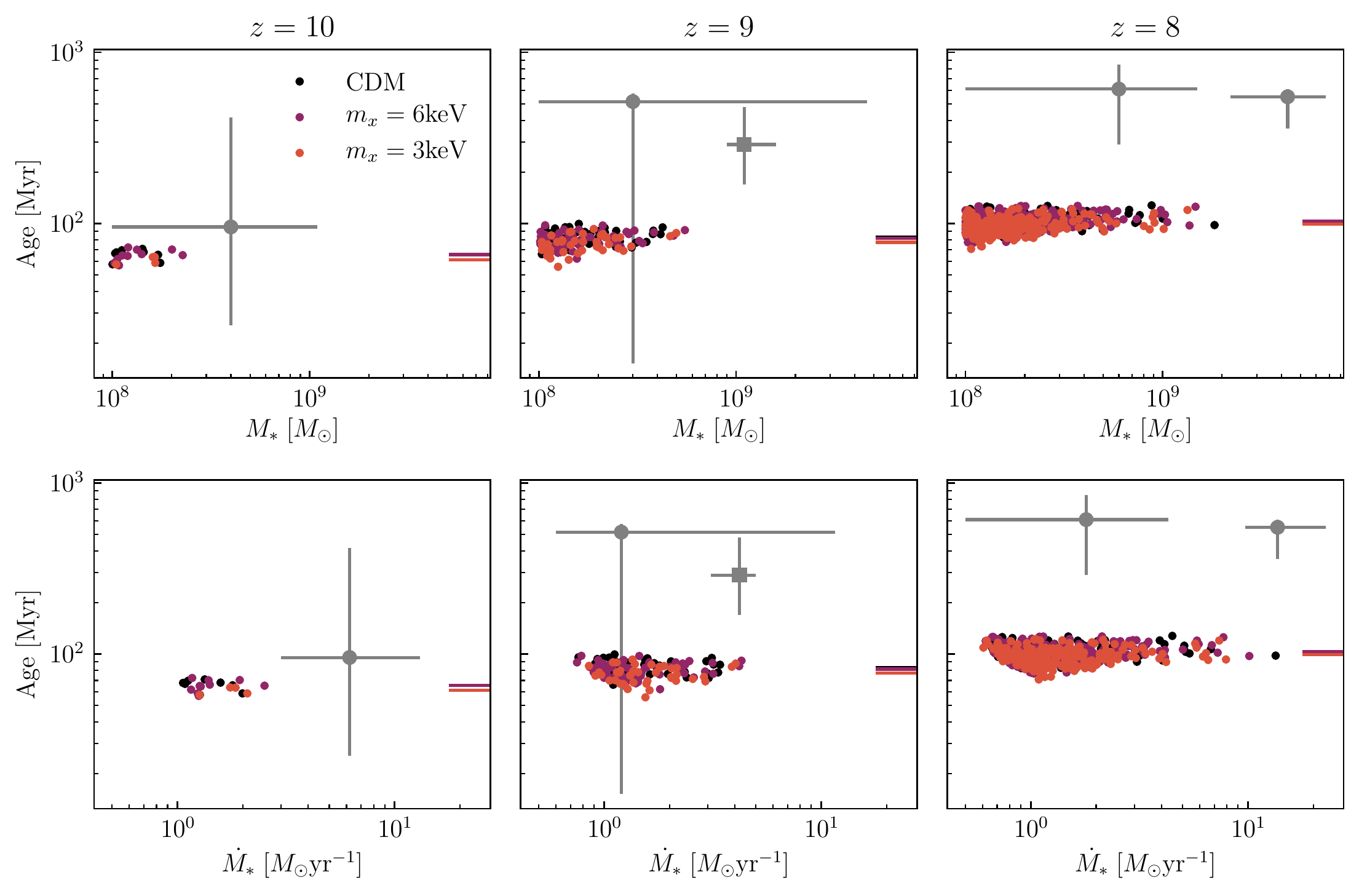}
    \caption{Comparison of ages, stellar masses, and star formation rates to observational data. Mass-weighted mean star particle age for each galaxy as a function of stellar mass (top row), and star-formation rate (bottom row) for each redshift bin analyzed (columns). Colors indicate different dark matter cosmologies as in Figure \ref{fig:age_CDFs}, and horizontal ticks indicate average ages as shown in Table~\ref{table:ages}. Gray points show observational data and errors quoted in \citet[][circles]{Strait2020} and \citet[][square]{Hashimoto2018}. The data are consistent with the predictions of all three simulations since they currently lack the precision to distinguish between different cosmologies.}
    \label{fig:galaxy_age_Ms_SFR}
\end{figure*}

Figure \ref{fig:galaxy_age_Ms_SFR} shows the ages of massive galaxies in CROC simulations with different dark matter cosmologies as a function of stellar mass and star formation rate, compared to observational constraints from \citet{Strait2020} and \citet{Hashimoto2018}. These constraints are obtained by forward-modeling the Spectral Energy Distribution (SED) of a stellar population to fit observed broadband magnitudes of these galaxies. Consequently, age estimates for these galaxies have much larger uncertainties than the few-Myr differences in average age between 3 keV WDM and CDM.

\subsubsection{Spectra and Balmer Break Distributions}

\begin{figure}
    \centering
    \includegraphics[width=\columnwidth]{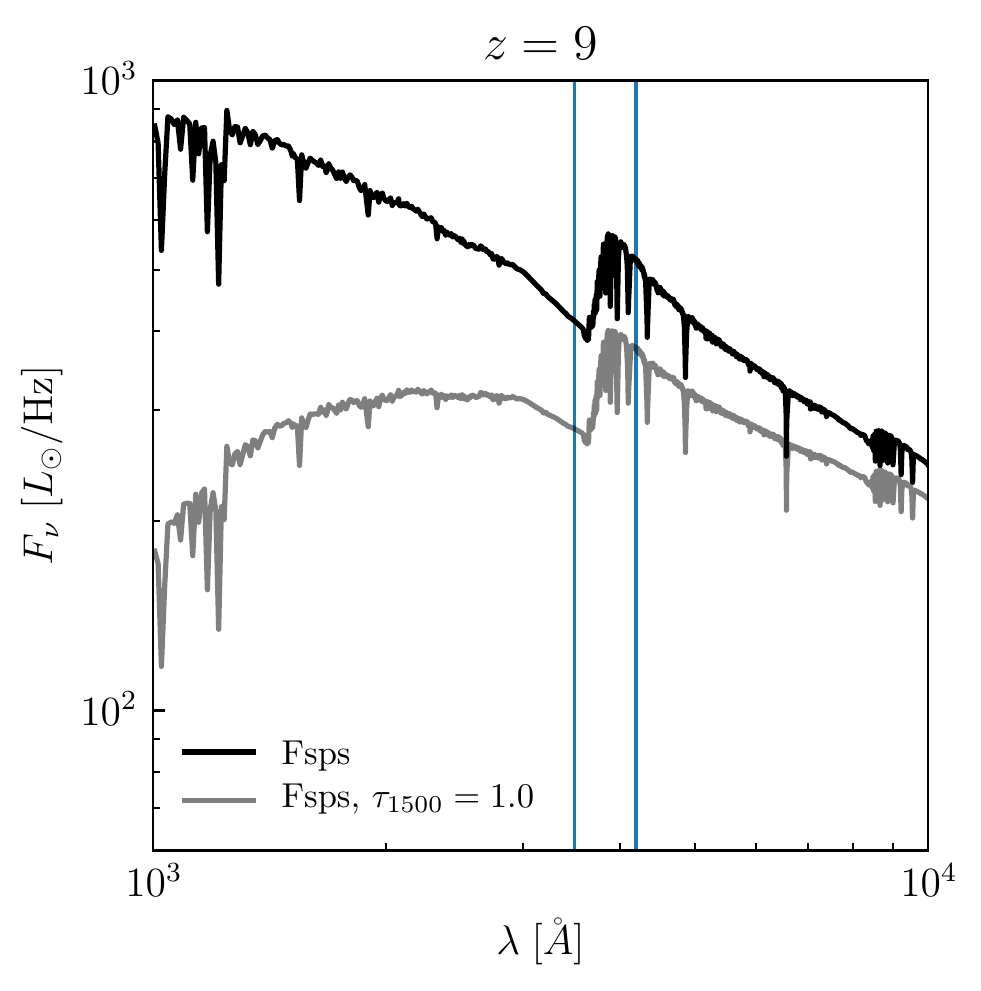}
    \caption{Example spectra. The Spectral Energy Distribution for a representative galaxy in the CDM simulation at $z=9$ ($M_{\rm vir} = 1.3\times 10^{11}M_{\odot}$, $M_* = 2.0\times 10^{8}M_{\odot}$, Age=$83.5$ Myr). The black line shows the unattenuated spectrum, calculated with the Flexible Stellar Population Synthesis code \citep{Conroy2009, Conroy2010, python_fsps}, while the gray line shows the spectrum assuming an optical depth due to dust at $\lambda = 1500\AA$ of $\tau_{1500} = 1.0$ and a power-law dust attenuation law with $\tau \propto \lambda^{-1.1}$. Vertical blue lines show $\lambda=3500\AA$ and $\lambda=4200\AA$, the wavelengths used to estimate Balmer break values shown in Fig.~\ref{fig:spec_analysis}.}
    \label{fig:spec_example}
\end{figure}

\begin{figure*}
    \centering
    \includegraphics[width=\textwidth]{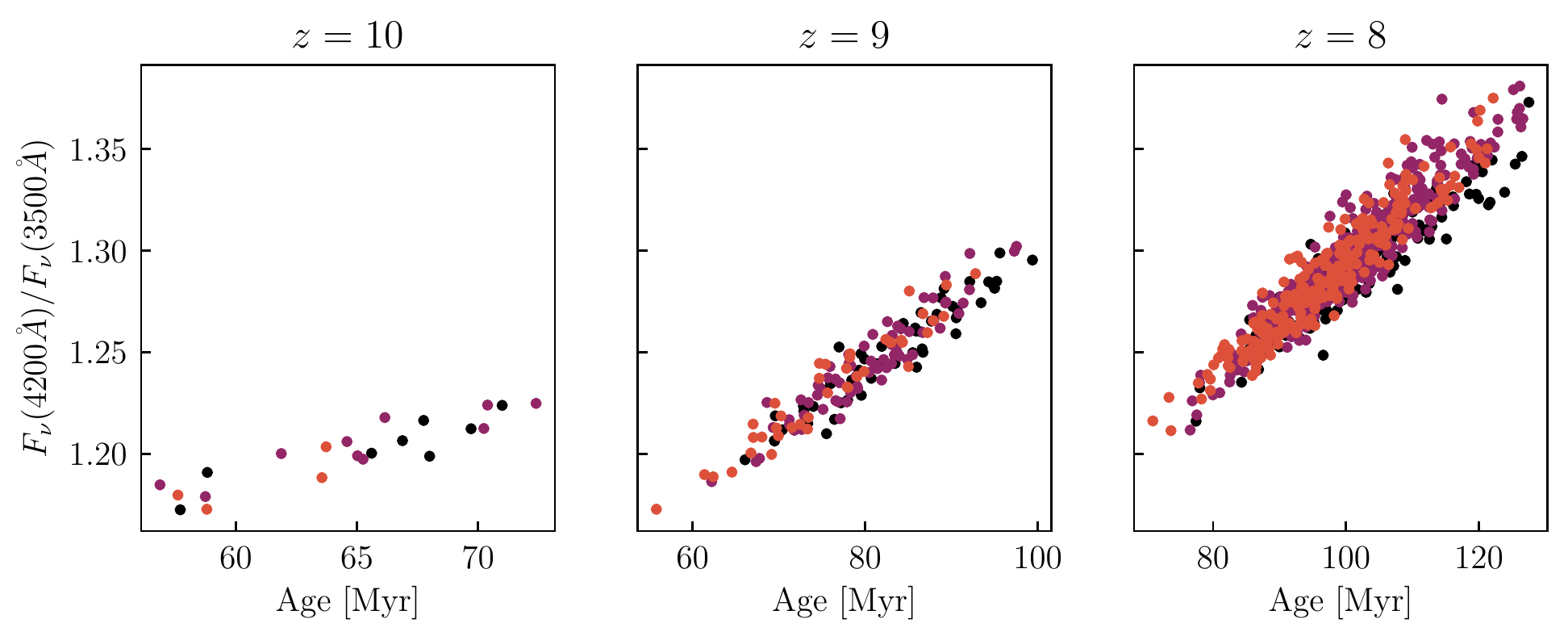}
    \caption{Balmer break distribution. Values of the Balmer break, quantified by the ratio of SED values at $\lambda = 4200\AA$ and $\lambda=3500\AA$, as a function of galaxy stellar population age.}
    \label{fig:spec_analysis}
\end{figure*}

Since the stellar population properties of unresolved galaxies are only observationally accessible through their effect on the total galaxy SED, we present predictions for the spectra of the massive galaxies in our simulations. We employ the Flexible Stellar Population Synthesis (FSPS) code \citep{Conroy2009, Conroy2010, python_fsps} in these calculations.  Figure~\ref{fig:spec_example} shows an example spectrum of a representative galaxy in the CDM simulation at $z=9$, with a halo mass of $M_{\rm vir} = 1.3\times 10^{11}M_{\odot}$, a stellar mass of $M_{*} = 2.0\times 10^{8}M_{\odot}$, and a mass-weighted mean stellar age of $83.5$ Myr. The spectrum is presented as both unattenuated (black line) and attenuated assuming an optical depth due to dust at $\lambda = 1500\AA$ of $\tau_{1500} = 1.0$ and a power-law dust attenuation law with $\tau \propto \lambda^{-1.1}$, \citep[i.e. an approximation of the SMC dust optical depth, see][]{WeingartnerDraine2001, Gnedin2008, CROCI}. \citet{KhakhalevaLi2016} estimated the optical depth due to dust in the CROC galaxies by post-processing the simulations with a radiative transfer code, and found values of $\tau_{1500}\sim0.2-0.6$ \citep[also assuming the SMC dust model from][]{WeingartnerDraine2001}. The spectra shown in Figure~\ref{fig:spec_example} therefore bracket the plausible range of attenuation in the simulations. 

Figure~\ref{fig:spec_analysis} shows the value of the Balmer break in the unattenuated spectrum of each galaxy, quantified as the ratio of SED flux (per frequency) at $4200$ and $3500\AA$. The value of this ratio is strongly correlated with the average age of the stellar population, as expected. The $\tau_{1500} = 1.0$ attenuation shown in Figure~\ref{fig:spec_example} would increase these by a factor of 1.07. 

\section{Discussion}

\subsection{Caveats}

\subsubsection{Convergence of Individual Galaxy Properties}
\label{sec:convergence}

\begin{figure}
    \centering
    \includegraphics[width=\columnwidth]{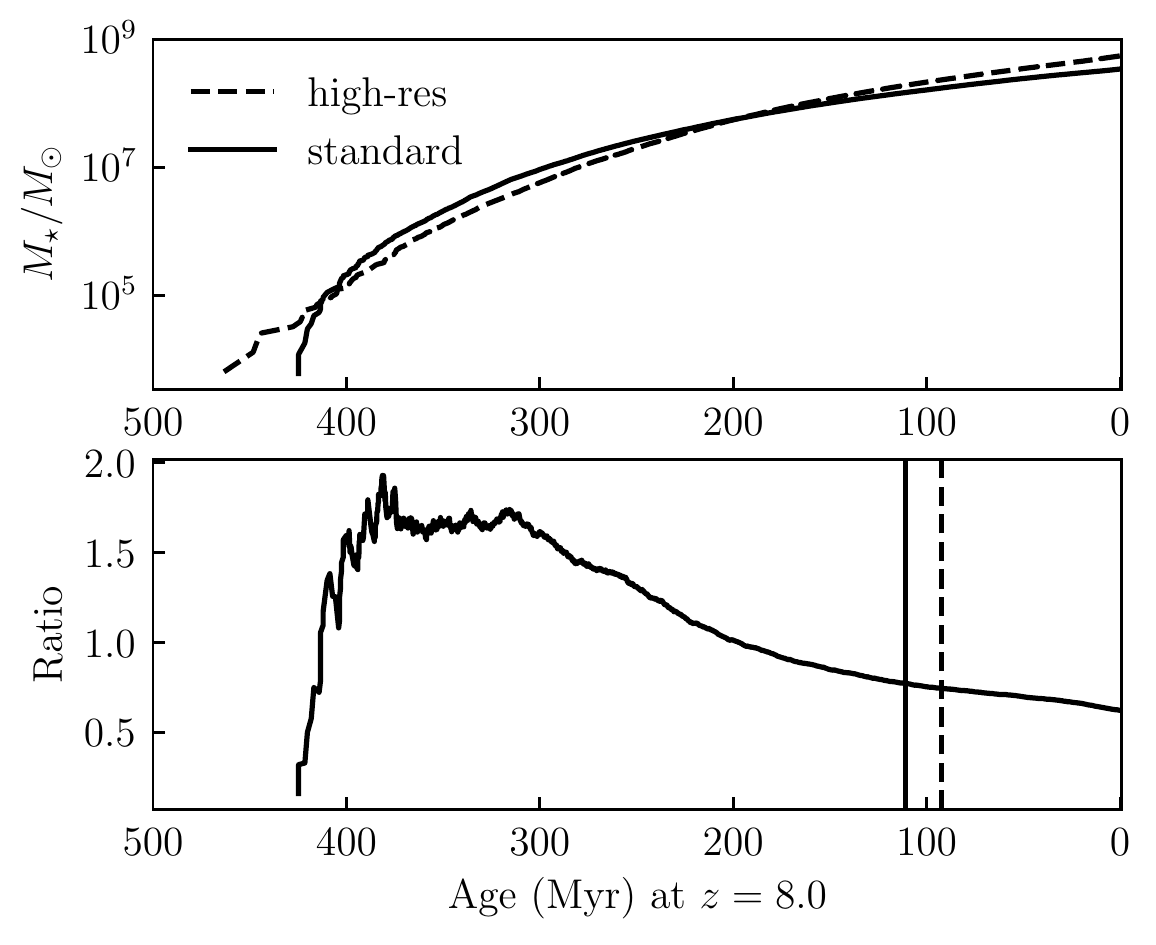}
    \caption{Dependence of galaxy star formation histories on simulation resolution. The top panel shows CDF of stellar particle ages for the most massive galaxy in the CDM simulation compared to the same galaxy in a higher-resolution re-simulation. The ratio of these CDFs and the mass-weighted mean age (vertical lines) for each galaxy are shown in the bottom panel.}
    \label{fig:res_test}
\end{figure}

 \citet{CROC_Convergence} explored the convergence properties of statistical measures of cosmological galaxy formation in the CROC simulations. In this section we assess the convergence of individual galaxy properties most relevant to the galaxy ages quoted above. Figure \ref{fig:res_test} compares the star particle age distribution of the most massive galaxy in the CDM simulations at standard and high resolution. All of the previous results were for simulations run at the standard resolution. The initial conditions of the high-resolution simulation have the same large-scale modes as those of the standard resolution ICs. There is therefore a one-to-one correspondence between well resolved individual galaxies from simulations with different resolutions. The same galaxy has a larger stellar mass by a factor of $\approx 1.5$ ($5.6\times10^8M_{\odot}$ versus $3.4\times10^8M_{\odot}$), and a younger mass-weighted mean age by $\sim 20\%$ (92 Myr vs. 111 Myr) in the high versus standard resolution simulations, respectively. This change is comparable to the differences in mean age between cosmologies, where they exist. We thus caution that the differences in stellar ages between different cosmologies quoted above may not be numerically converged. 
 
\subsubsection{Star Formation Model}

While the uncertainties in current observations are sufficiently large that the data are consistent with all simulated cosmologies, the two galaxies in each of the the $z = 9$ and $z = 8$ redshift bins are systematically older than those in the CROC simulations. It is perhaps interesting to note that, despite these higher most-likely ages, the most-likely stellar masses and star formation rates of observed galaxies agree well with the most massive simulated galaxies at these redshifts. This implies that it may be possible to form such massive galaxies by these redshifts without the early star formation claimed in these observational analyses.

Alternatively, if future observational data confirm the presence of such early-forming stellar populations, this would be a strong indication of star formation in atomic gas. None of the simulations we analyze form stars earlier than 150 Myr after the Big Bang, since CROC only allows star formation in molecular gas,  which has not yet formed by these cosmological times. Since the resolution of the CROC simulations was explicitly chosen to resolve all halos in which molecular gas can form, this is a robust prediction, and contradictory observational evidence would be a compelling indication of star formation in a fundamentally different physical regime than observed in the local universe. However, conclusive results on any of these topics await significantly more observational data.

\subsection{Comparison to Previous Studies}

\citet{Binggeli2019} calculated the Balmer break distributions of high-redshift ($z > 7$) galaxies from three other state-of-the-art simulation suites: FIRE-2 \citep{Ma2018}, FirstLight \citep{Ceverino2017}, and \citet{Shimizu2016}. They accounted for nebular emission and explored several choices for including the effect of dust attenuation, all of which broadly predicted typical Balmer breaks of 1.2-1.5 for the galaxies in these simulations, consistent with our results. We note that their predictions appear to span a wider range of values than ours. This may reflect the different prescriptions for star formation and feedback - and hence the age distribution of stellar populations - in the different simulations, or their choices for simulation post-processing, some of which account for radiative transfer effects that we do not explore. We also note that they employ different stellar spectral synthesis templates - those presented in \citet{Zachrisson2011}, which are based on STARBURST99 \citep{Starburst99}. We have checked that replacing FSPS with the \citet{Zachrisson2011} tables in our analysis systematically increases the predicted Balmer break value by $\sim 0.1$, keeping our quantitative results broadly consistent with theirs. 

The results of \citet{Katz2019} provide a potential resolution to the inability of these simulations to reproduce the \citet{Hashimoto2018} observations: they find that the inhomogeneous distribution of dust in the ISM of their simulations \citep{Rosdahl2018} preferentially attenuates the light from young, UV-bright stellar populations, thereby increasing the Balmer break without the need for a significant evolved stellar population. It is therefore possible that the CROC simulations, as well as those analyzed in \citet{Binggeli2019}, have realistic star formation histories, but need even more dust attenuation. Given the crucial importance of dust for observationally constraining the star formation histories of high-redshift galaxies, the large uncertainties in dust properties and dynamics in the high-redshift ISM demand further theoretical investigation.

Furthermore, several previous studies have investigated the effect of WDM cosmologies on the star formation histories of galaxies, although primarily at different redshifts, galaxy masses, and dark matter particle masses than we investigate here, complicating direct comparison. Nonetheless, \citet{Dayal2015}, also explored early galaxy formation in WDM cosmologies using a semianalytic model tuned to matched the high-redshift UV luminosity function. They nominally found even smaller differences between the galaxy formation histories in different cosmologies than we do, only identifying significant differences for WDM thermal relic particle masses of $m_x \le 2{\rm keV}$. However, the differences between methods of comparison - namely the statistical properties of the galaxy population and the galaxy masses analyzed - preclude a strong statement on the lack of consistency with our results.

Conversely \citet{Lovell2020} compared simulated Local Group analogs with the EAGLE galaxy formation model in CDM, WDM, and Self-Interacting Dark Matter, and report a larger ($\sim 100-200$ Myr) delay in galaxy formation between CDM and WDM than we find. However, this likely is not a fair comparison because they analyze $\sim 10^{12}M_{\odot}$ halos at $z=0$, while the halos in our simulations are the progenitors of massive cluster galaxies. Moreover, the ``condensation time'' (the time at which a halo is first able to form stars based on a mass threshold) they use to quantify the differences between the different cosmologies is certainly more sensitive to the dark matter power spectrum than the average stellar age. Finally, they too note a significant resolution dependence on the final stellar masses of their simulated galaxies, highlighting the room for progress toward numerically robust predictions on this topic.  

\section{Conclusions}

\begin{enumerate}
    \item The mean stellar ages of massive galaxies ($M_* > 10^8M_{\odot}$) at high redshift ($z \ge 8$) in the CROC simulations are significantly different, by $\sim 5$ Myr, between CDM and $m_x = 3$keV WDM cosmologies, even when the galaxy formation parameters are re-tuned to match existing observational constraints equally well. Galaxy stellar ages in CDM and $m_x = 6$keV WDM are indistinguishable at these masses and redshifts.
    \item Current constraints on the stellar populations of high-redshift galaxies are insufficiently precise to distinguish between the predictions of different cosmologies.
    \item The spectra of these simulated galaxies fail to produce a large Balmer break, but this prediction is likely sensitive to the deeply uncertain properties of the dust. 
    \item The masses and ages of the galaxies we analyze probably are not converged to the precision needed for the predictions for differences between cosmologies to be numerically robust. 
\end{enumerate}

\acknowledgments
This manuscript has been authored by Fermi Research Alliance, LLC, under Contract No. DE-AC02-07CH11359 with the U.S. Department of Energy, Office of Science, Office of High Energy Physics. It is based upon work that is supported by the Visiting Scholars Award Program of the Universities Research Association at Fermilab. Our analysis made use of the following publicly available software packages: Matplotlib \citep{Matplotlib}, SciPy \citep{SciPy}, NumPy \citep{NumPy}, COLOSSUS \citep{COLOSSUS}, and yt \citep{yt}.


\appendix

\section{$6\;{\rm keV}$ Calibration}
\label{sec:appendix_calibration}

\begin{figure}
    \centering
    \includegraphics[width=0.6\columnwidth]{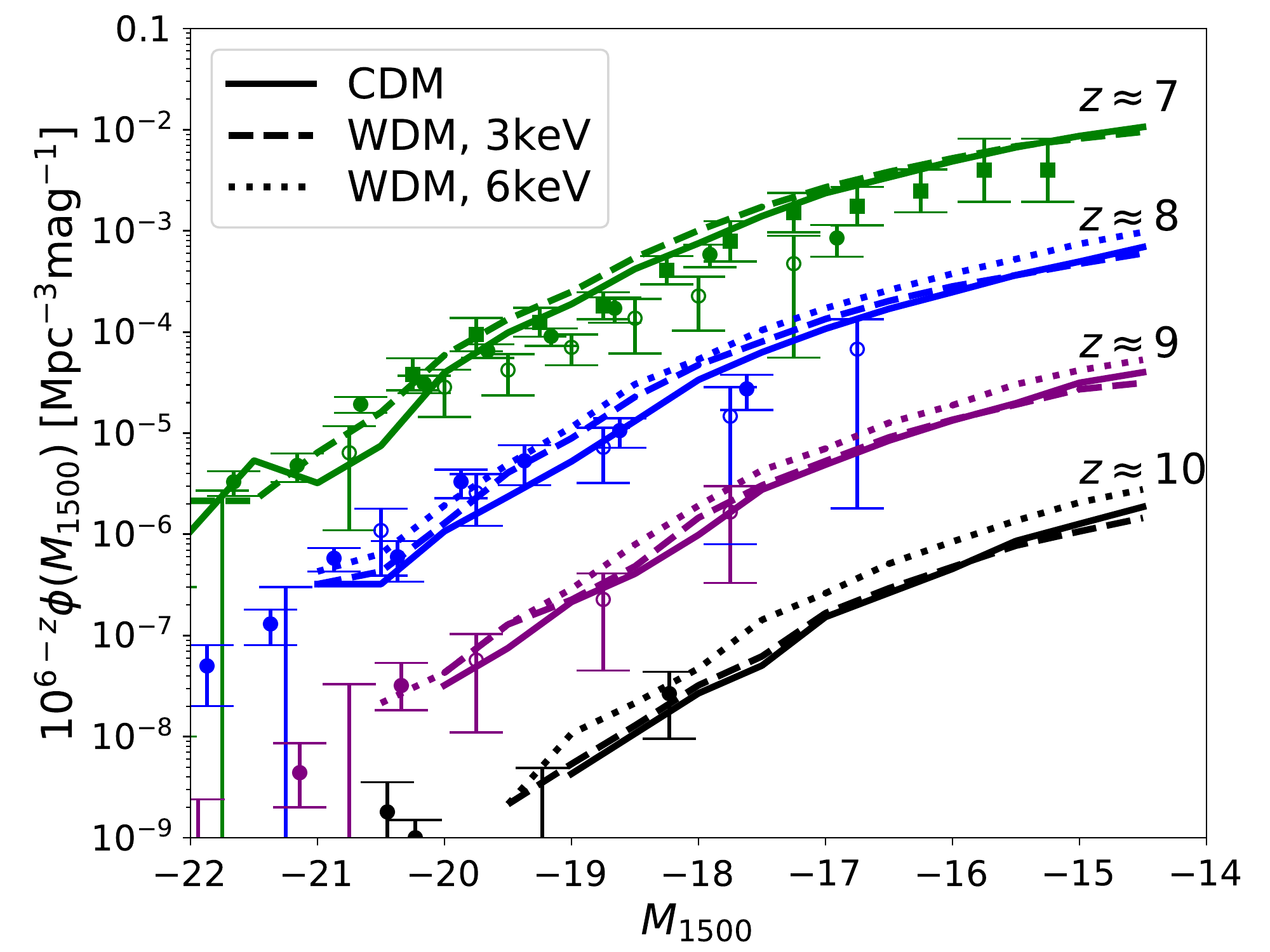}
    \caption{UV luminosity functions. Comparison of observed high-redshift luminosity functions and predictions from the WDM simulations presented in this paper. Data are from \citet[][unfilled circles]{Laporte2016ApJ...820...98L}, \citet[][filled circles]{Bouwens2015ApJ...803...34B, Bouwens2016ApJ...830...67B, Bouwens2017ApJ...843..129B}, and \citet[][squares]{Atek2015ApJ...814...69A}.}
    \label{fig:lf-wdm}
\end{figure}

In this paper we present a new simulation to model $6 {\rm keV}$ WDM. Due to computational constraints we could not recalibrate the star formation prescription for this simulation, so the same parameters as $3 {\rm keV}$ WDM were adopted. To assess the extent to which is simulation is out of calibration, Figure~\ref{fig:lf-wdm} shows the UV luminosity functions of this simulation compared to the other cosmologies and data.

\section{Stellar Mass-Halo Mass Relation}
\label{sec:appendix_msmh}
\begin{figure}
    \centering
    \includegraphics{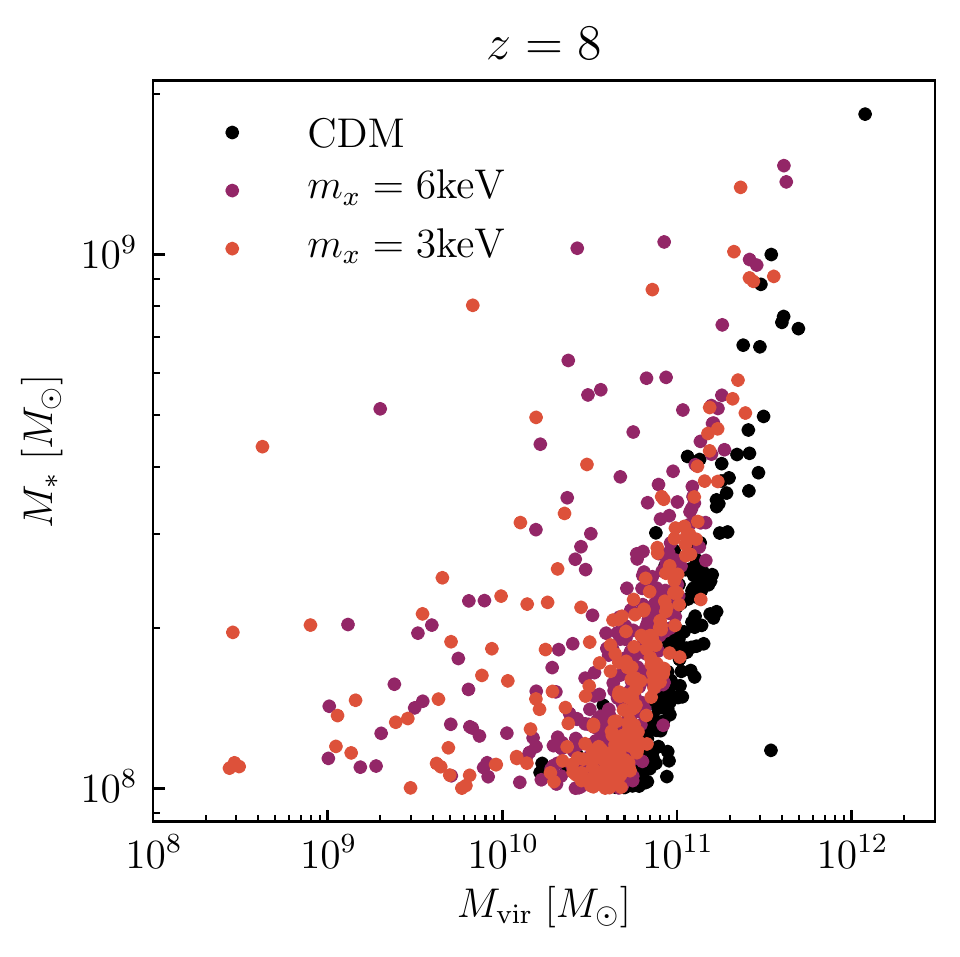}
    \caption{$z=8$ stellar mass-halo mass relation in CDM and WDM cosmologies. }
    \label{fig:msmh}
\end{figure}

To demonstrate the magnitude of predicted differences in the galaxy-halo connection and therefore galaxy bias between CDM and WDM, we show the stellar mass-halo mass relation for our simulations at $z=8$ in Figure~\ref{fig:msmh}. The WDM relation at both masses is slightly offset from that of CDM, suggesting that this difference is due primarily to the different depletion timescales used in the star formation models (see Section~\ref{sec:calibration}).

\bibliography{bibliography} 

\begin{thebibliography}{}
\expandafter\ifx\csname natexlab\endcsname\relax\def\natexlab#1{#1}\fi
\providecommand{\url}[1]{\href{#1}{#1}}

\bibitem[{{Alam} {et~al.}(2017){Alam}, {Ata}, {Bailey}, {Beutler}, {Bizyaev},
  {Blazek}, {Bolton}, {Brownstein}, {Burden}, {Chuang}, {Comparat}, {Cuesta},
  {Dawson}, {Eisenstein}, {Escoffier}, {Gil-Mar{\'\i}n}, {Grieb}, {Hand}, {Ho},
  {Kinemuchi}, {Kirkby}, {Kitaura}, {Malanushenko}, {Malanushenko}, {Maraston},
  {McBride}, {Nichol}, {Olmstead}, {Oravetz}, {Padmanabhan},
  {Palanque-Delabrouille}, {Pan}, {Pellejero-Ibanez}, {Percival}, {Petitjean},
  {Prada}, {Price-Whelan}, {Reid}, {Rodr{\'\i}guez-Torres}, {Roe}, {Ross},
  {Ross}, {Rossi}, {Rubi{\~n}o-Mart{\'\i}n}, {Saito}, {Salazar-Albornoz},
  {Samushia}, {S{\'a}nchez}, {Satpathy}, {Schlegel}, {Schneider},
  {Sc{\'o}ccola}, {Seo}, {Sheldon}, {Simmons}, {Slosar}, {Strauss}, {Swanson},
  {Thomas}, {Tinker}, {Tojeiro}, {Maga{\~n}a}, {Vazquez}, {Verde}, {Wake},
  {Wang}, {Weinberg}, {White}, {Wood-Vasey}, {Y{\`e}che}, {Zehavi}, {Zhai}, \&
  {Zhao}}]{SDSSDR12Cosmo}
{Alam}, S., {Ata}, M., {Bailey}, S., {et~al.} 2017, \mnras, 470, 2617

\bibitem[{{Atek} {et~al.}(2015){Atek}, {Richard}, {Jauzac}, {Kneib},
  {Natarajan}, {Limousin}, {Schaerer}, {Jullo}, {Ebeling}, {Egami}, \&
  {Clement}}]{Atek2015ApJ...814...69A}
{Atek}, H., {Richard}, J., {Jauzac}, M., {et~al.} 2015, \apj, 814, 69

\bibitem[{{Behroozi} {et~al.}(2013){Behroozi}, {Wechsler}, \& {Wu}}]{ROCKSTAR}
{Behroozi}, P.~S., {Wechsler}, R.~H., \& {Wu}, H.-Y. 2013, \apj, 762, 109

\bibitem[{{Binggeli} {et~al.}(2019){Binggeli}, {Zackrisson}, {Ma}, {Inoue},
  {Vikaeus}, {Hashimoto}, {Mawatari}, {Shimizu}, \& {Ceverino}}]{Binggeli2019}
{Binggeli}, C., {Zackrisson}, E., {Ma}, X., {et~al.} 2019, \mnras, 489, 3827

\bibitem[{{Bouwens} {et~al.}(2017){Bouwens}, {Oesch}, {Illingworth}, {Ellis},
  \& {Stefanon}}]{Bouwens2017ApJ...843..129B}
{Bouwens}, R.~J., {Oesch}, P.~A., {Illingworth}, G.~D., {Ellis}, R.~S., \&
  {Stefanon}, M. 2017, \apj, 843, 129

\bibitem[{{Bouwens} {et~al.}(2015){Bouwens}, {Illingworth}, {Oesch}, {Trenti},
  {Labb{\'e}}, {Bradley}, {Carollo}, {van Dokkum}, {Gonzalez}, {Holwerda},
  {Franx}, {Spitler}, {Smit}, \& {Magee}}]{Bouwens2015ApJ...803...34B}
{Bouwens}, R.~J., {Illingworth}, G.~D., {Oesch}, P.~A., {et~al.} 2015, \apj,
  803, 34

\bibitem[{{Bouwens} {et~al.}(2016){Bouwens}, {Oesch}, {Labb{\'e}},
  {Illingworth}, {Fazio}, {Coe}, {Holwerda}, {Smit}, {Stefanon}, {van Dokkum},
  {Trenti}, {Ashby}, {Huang}, {Spitler}, {Straatman}, {Bradley}, \&
  {Magee}}]{Bouwens2016ApJ...830...67B}
{Bouwens}, R.~J., {Oesch}, P.~A., {Labb{\'e}}, I., {et~al.} 2016, \apj, 830, 67

\bibitem[{{Carucci} {et~al.}(2015){Carucci}, {Villaescusa-Navarro}, {Viel}, \&
  {Lapi}}]{Carucci2015}
{Carucci}, I.~P., {Villaescusa-Navarro}, F., {Viel}, M., \& {Lapi}, A. 2015,
  \jcap, 2015, 047

\bibitem[{{Ceverino} {et~al.}(2017){Ceverino}, {Glover}, \&
  {Klessen}}]{Ceverino2017}
{Ceverino}, D., {Glover}, S. C.~O., \& {Klessen}, R.~S. 2017, \mnras, 470, 2791

\bibitem[{{Conroy} \& {Gunn}(2010)}]{Conroy2010}
{Conroy}, C., \& {Gunn}, J.~E. 2010, \apj, 712, 833

\bibitem[{{Conroy} {et~al.}(2009){Conroy}, {Gunn}, \& {White}}]{Conroy2009}
{Conroy}, C., {Gunn}, J.~E., \& {White}, M. 2009, \apj, 699, 486

\bibitem[{{Dayal} {et~al.}(2015){Dayal}, {Mesinger}, \& {Pacucci}}]{Dayal2015}
{Dayal}, P., {Mesinger}, A., \& {Pacucci}, F. 2015, \apj, 806, 67

\bibitem[{{Diemer}(2018)}]{COLOSSUS}
{Diemer}, B. 2018, \apjs, 239, 35

\bibitem[{{Eisenstein} \& {Hut}(1998)}]{HOP}
{Eisenstein}, D.~J., \& {Hut}, P. 1998, \apj, 498, 137

\bibitem[{Foreman-Mackey {et~al.}(2014)Foreman-Mackey, Sick, \&
  Johnson}]{python_fsps}
Foreman-Mackey, D., Sick, J., \& Johnson, B. 2014, python-fsps: Python bindings
  to FSPS (v0.1.1), vv0.1.1,  Zenodo, doi:10.5281/zenodo.12157.
\newblock \url{https://doi.org/10.5281/zenodo.12157}

\bibitem[{{Gilman} {et~al.}(2020){Gilman}, {Birrer}, {Nierenberg}, {Treu},
  {Du}, \& {Benson}}]{Gilman2019}
{Gilman}, D., {Birrer}, S., {Nierenberg}, A., {et~al.} 2020, \mnras, 491, 6077

\bibitem[{{Gnedin}(2014)}]{CROCI}
{Gnedin}, N.~Y. 2014, \apj, 793, 29

\bibitem[{{Gnedin}(2016)}]{CROC_Convergence}
---. 2016, \apj, 821, 50

\bibitem[{{Gnedin} {et~al.}(2017){Gnedin}, {Becker}, \& {Fan}}]{Gnedin2017}
{Gnedin}, N.~Y., {Becker}, G.~D., \& {Fan}, X. 2017, \apj, 841, 26

\bibitem[{{Gnedin} \& {Kaurov}(2014)}]{CROCII}
{Gnedin}, N.~Y., \& {Kaurov}, A.~A. 2014, \apj, 793, 30

\bibitem[{{Gnedin} {et~al.}(2008){Gnedin}, {Kravtsov}, \& {Chen}}]{Gnedin2008}
{Gnedin}, N.~Y., {Kravtsov}, A.~V., \& {Chen}, H.-W. 2008, \apj, 672, 765

\bibitem[{{Gnedin} {et~al.}(2011){Gnedin}, {Kravtsov}, \&
  {Rudd}}]{Gnedin2011_DC}
{Gnedin}, N.~Y., {Kravtsov}, A.~V., \& {Rudd}, D.~H. 2011, \apjs, 194, 46

\bibitem[{{Hashimoto} {et~al.}(2018){Hashimoto}, {Laporte}, {Mawatari},
  {Ellis}, {Inoue}, {Zackrisson}, {Roberts-Borsani}, {Zheng}, {Tamura},
  {Bauer}, {Fletcher}, {Harikane}, {Hatsukade}, {Hayatsu}, {Matsuda}, {Matsuo},
  {Okamoto}, {Ouchi}, {Pell{\'o}}, {Rydberg}, {Shimizu}, {Taniguchi},
  {Umehata}, \& {Yoshida}}]{Hashimoto2018}
{Hashimoto}, T., {Laporte}, N., {Mawatari}, K., {et~al.} 2018, Nature, 557, 392

\bibitem[{Hunter(2007)}]{Matplotlib}
Hunter, J.~D. 2007, Computing in Science \& Engineering, 9, 90

\bibitem[{{Ir{\v{s}}i{\v{c}}} {et~al.}(2017){Ir{\v{s}}i{\v{c}}}, {Viel},
  {Haehnelt}, {Bolton}, {Cristiani}, {Becker}, {D'Odorico}, {Cupani}, {Kim},
  {Berg}, {L{\'o}pez}, {Ellison}, {Christensen}, {Denney}, \&
  {Worseck}}]{Irsivic2017}
{Ir{\v{s}}i{\v{c}}}, V., {Viel}, M., {Haehnelt}, M.~G., {et~al.} 2017, \prd,
  96, 023522

\bibitem[{Jones {et~al.}(2001--)Jones, Oliphant, Peterson, {et~al.}}]{SciPy}
Jones, E., Oliphant, T., Peterson, P., {et~al.} 2001--, {SciPy}: Open source
  scientific tools for {Python}, , .
\newblock \url{http://www.scipy.org/}

\bibitem[{{Kang}(2020)}]{Kang2020}
{Kang}, X. 2020, \mnras, 491, 2520

\bibitem[{{Kang} {et~al.}(2013){Kang}, {Macci{\`o}}, \& {Dutton}}]{Kang2013}
{Kang}, X., {Macci{\`o}}, A.~V., \& {Dutton}, A.~A. 2013, \apj, 767, 22

\bibitem[{{Katz} {et~al.}(2019){Katz}, {Laporte}, {Ellis}, {Devriendt}, \&
  {Slyz}}]{Katz2019}
{Katz}, H., {Laporte}, N., {Ellis}, R.~S., {Devriendt}, J., \& {Slyz}, A. 2019,
  \mnras, 484, 4054

\bibitem[{{Kennedy} {et~al.}(2014){Kennedy}, {Frenk}, {Cole}, \&
  {Benson}}]{Kennedy2014}
{Kennedy}, R., {Frenk}, C., {Cole}, S., \& {Benson}, A. 2014, \mnras, 442, 2487

\bibitem[{{Khakhaleva-Li} \& {Gnedin}(2016)}]{KhakhalevaLi2016}
{Khakhaleva-Li}, Z., \& {Gnedin}, N.~Y. 2016, \apj, 820, 133

\bibitem[{{Kravtsov}(1999)}]{Kravtsov1999PhDT}
{Kravtsov}, A.~V. 1999, PhD thesis, NEW MEXICO STATE UNIVERSITY

\bibitem[{{Kravtsov} {et~al.}(2002){Kravtsov}, {Klypin}, \&
  {Hoffman}}]{Kravtsov2002}
{Kravtsov}, A.~V., {Klypin}, A., \& {Hoffman}, Y. 2002, \apj, 571, 563

\bibitem[{{Laporte} {et~al.}(2016){Laporte}, {Infante}, {Troncoso Iribarren},
  {Zheng}, {Molino}, {Bauer}, {Bina}, {Broadhurst}, {Chilingarian}, {Huang},
  {Garcia}, {Kim}, {Marques-Chaves}, {Moustakas}, {Pell{\'o}},
  {P{\'e}rez-Fournon}, {Shu}, {Streblyanska}, \&
  {Zitrin}}]{Laporte2016ApJ...820...98L}
{Laporte}, N., {Infante}, L., {Troncoso Iribarren}, P., {et~al.} 2016, \apj,
  820, 98

\bibitem[{{Leitherer} {et~al.}(1999){Leitherer}, {Schaerer}, {Goldader},
  {Delgado}, {Robert}, {Kune}, {de Mello}, {Devost}, \&
  {Heckman}}]{Starburst99}
{Leitherer}, C., {Schaerer}, D., {Goldader}, J.~D., {et~al.} 1999, \apjs, 123,
  3

\bibitem[{{Lovell} {et~al.}(2020){Lovell}, {Hellwing}, {Ludlow}, {Zavala},
  {Robertson}, {Fattahi}, {Frenk}, \& {Hardwick}}]{Lovell2020}
{Lovell}, M.~R., {Hellwing}, W., {Ludlow}, A., {et~al.} 2020, \mnras, 498, 702

\bibitem[{{Ma} {et~al.}(2018){Ma}, {Hopkins}, {Garrison-Kimmel},
  {Faucher-Gigu{\`e}re}, {Quataert}, {Boylan-Kolchin}, {Hayward}, {Feldmann},
  \& {Kere{\v{s}}}}]{Ma2018}
{Ma}, X., {Hopkins}, P.~F., {Garrison-Kimmel}, S., {et~al.} 2018, \mnras, 478,
  1694

\bibitem[{{Menci} {et~al.}(2016){Menci}, {Grazian}, {Castellano}, \&
  {Sanchez}}]{Menci2016}
{Menci}, N., {Grazian}, A., {Castellano}, M., \& {Sanchez}, N.~G. 2016, \apjl,
  825, L1

\bibitem[{{Naab} \& {Ostriker}(2017)}]{NaabOstrikerARAA2017}
{Naab}, T., \& {Ostriker}, J.~P. 2017, \araa, 55, 59

\bibitem[{{Planck Collaboration} {et~al.}(2020){Planck Collaboration},
  {Aghanim}, {Akrami}, {Ashdown}, {Aumont}, {Baccigalupi}, {Ballardini},
  {Banday}, {Barreiro}, {Bartolo}, {Basak}, {Battye}, {Benabed}, {Bernard},
  {Bersanelli}, {Bielewicz}, {Bock}, {Bond}, {Borrill}, {Bouchet}, {Boulanger},
  {Bucher}, {Burigana}, {Butler}, {Calabrese}, {Cardoso}, {Carron},
  {Challinor}, {Chiang}, {Chluba}, {Colombo}, {Combet}, {Contreras}, {Crill},
  {Cuttaia}, {de Bernardis}, {de Zotti}, {Delabrouille}, {Delouis}, {Di
  Valentino}, {Diego}, {Dor{\'e}}, {Douspis}, {Ducout}, {Dupac}, {Dusini},
  {Efstathiou}, {Elsner}, {En{\ss}lin}, {Eriksen}, {Fantaye}, {Farhang},
  {Fergusson}, {Fernandez-Cobos}, {Finelli}, {Forastieri}, {Frailis},
  {Fraisse}, {Franceschi}, {Frolov}, {Galeotta}, {Galli}, {Ganga},
  {G{\'e}nova-Santos}, {Gerbino}, {Ghosh}, {Gonz{\'a}lez-Nuevo}, {G{\'o}rski},
  {Gratton}, {Gruppuso}, {Gudmundsson}, {Hamann}, {Handley}, {Hansen},
  {Herranz}, {Hildebrandt}, {Hivon}, {Huang}, {Jaffe}, {Jones}, {Karakci},
  {Keih{\"a}nen}, {Keskitalo}, {Kiiveri}, {Kim}, {Kisner}, {Knox},
  {Krachmalnicoff}, {Kunz}, {Kurki-Suonio}, {Lagache}, {Lamarre}, {Lasenby},
  {Lattanzi}, {Lawrence}, {Le Jeune}, {Lemos}, {Lesgourgues}, {Levrier},
  {Lewis}, {Liguori}, {Lilje}, {Lilley}, {Lindholm}, {L{\'o}pez-Caniego},
  {Lubin}, {Ma}, {Mac{\'\i}as-P{\'e}rez}, {Maggio}, {Maino}, {Mandolesi},
  {Mangilli}, {Marcos-Caballero}, {Maris}, {Martin}, {Martinelli},
  {Mart{\'\i}nez-Gonz{\'a}lez}, {Matarrese}, {Mauri}, {McEwen}, {Meinhold},
  {Melchiorri}, {Mennella}, {Migliaccio}, {Millea}, {Mitra},
  {Miville-Desch{\^e}nes}, {Molinari}, {Montier}, {Morgante}, {Moss}, {Natoli},
  {N{\o}rgaard-Nielsen}, {Pagano}, {Paoletti}, {Partridge}, {Patanchon},
  {Peiris}, {Perrotta}, {Pettorino}, {Piacentini}, {Polastri}, {Polenta},
  {Puget}, {Rachen}, {Reinecke}, {Remazeilles}, {Renzi}, {Rocha}, {Rosset},
  {Roudier}, {Rubi{\~n}o-Mart{\'\i}n}, {Ruiz-Granados}, {Salvati}, {Sandri},
  {Savelainen}, {Scott}, {Shellard}, {Sirignano}, {Sirri}, {Spencer},
  {Sunyaev}, {Suur-Uski}, {Tauber}, {Tavagnacco}, {Tenti}, {Toffolatti},
  {Tomasi}, {Trombetti}, {Valenziano}, {Valiviita}, {Van Tent}, {Vibert},
  {Vielva}, {Villa}, {Vittorio}, {Wandelt}, {Wehus}, {White}, {White},
  {Zacchei}, \& {Zonca}}]{Planck2018VI}
{Planck Collaboration}, {Aghanim}, N., {Akrami}, Y., {et~al.} 2020, \aap, 641,
  A6

\bibitem[{{Rosdahl} {et~al.}(2018){Rosdahl}, {Katz}, {Blaizot}, {Kimm},
  {Michel-Dansac}, {Garel}, {Haehnelt}, {Ocvirk}, \& {Teyssier}}]{Rosdahl2018}
{Rosdahl}, J., {Katz}, H., {Blaizot}, J., {et~al.} 2018, \mnras, 479, 994

\bibitem[{{Rudd} {et~al.}(2008){Rudd}, {Zentner}, \& {Kravtsov}}]{Rudd2008}
{Rudd}, D.~H., {Zentner}, A.~R., \& {Kravtsov}, A.~V. 2008, \apj, 672, 19

\bibitem[{{Shimizu} {et~al.}(2016){Shimizu}, {Inoue}, {Okamoto}, \&
  {Yoshida}}]{Shimizu2016}
{Shimizu}, I., {Inoue}, A.~K., {Okamoto}, T., \& {Yoshida}, N. 2016, \mnras,
  461, 3563

\bibitem[{{Strait} {et~al.}(2020){Strait}, {Brada{\v{c}}}, {Coe}, {Bradley},
  {Salmon}, {Lemaux}, {Huang}, {Zitrin}, {Sharon}, {Acebron}, {Andrade-Santos},
  {Avila}, {Frye}, {Hoag}, {Mahler}, {Nonino}, {Ogaz}, {Oguri}, {Ouchi},
  {Paterno-Mahler}, \& {Pelliccia}}]{Strait2020}
{Strait}, V., {Brada{\v{c}}}, M., {Coe}, D., {et~al.} 2020, \apj, 888, 124

\bibitem[{{Turk} {et~al.}(2011){Turk}, {Smith}, {Oishi}, {Skory}, {Skillman},
  {Abel}, \& {Norman}}]{yt}
{Turk}, M.~J., {Smith}, B.~D., {Oishi}, J.~S., {et~al.} 2011, The Astrophysical
  Journal Supplement Series, 192, 9

\bibitem[{Viel {et~al.}(2005)Viel, Lesgourgues, Haehnelt, Matarrese, \&
  Riotto}]{Viel2005}
Viel, M., Lesgourgues, J., Haehnelt, M.~G., Matarrese, S., \& Riotto, A. 2005,
  Phys. Rev. D, 71, 063534.
\newblock \url{https://link.aps.org/doi/10.1103/PhysRevD.71.063534}

\bibitem[{{Villanueva-Domingo} {et~al.}(2018){Villanueva-Domingo}, {Gnedin}, \&
  {Mena}}]{CROC_WDM}
{Villanueva-Domingo}, P., {Gnedin}, N.~Y., \& {Mena}, O. 2018, \apj, 852, 139

\bibitem[{Walt {et~al.}(2011)Walt, Colbert, \& Varoquaux}]{NumPy}
Walt, S. v.~d., Colbert, S.~C., \& Varoquaux, G. 2011, Computing in Science \&
  Engineering, 13, 22.
\newblock \url{https://aip.scitation.org/doi/abs/10.1109/MCSE.2011.37}

\bibitem[{{Weingartner} \& {Draine}(2001)}]{WeingartnerDraine2001}
{Weingartner}, J.~C., \& {Draine}, B.~T. 2001, \apj, 548, 296

\bibitem[{{Y{\`e}che} {et~al.}(2017){Y{\`e}che}, {Palanque-Delabrouille},
  {Baur}, \& {du Mas des Bourboux}}]{Yeche2017}
{Y{\`e}che}, C., {Palanque-Delabrouille}, N., {Baur}, J., \& {du Mas des
  Bourboux}, H. 2017, \jcap, 2017, 047

\bibitem[{{Zackrisson} {et~al.}(2011){Zackrisson}, {Rydberg}, {Schaerer},
  {{\"O}stlin}, \& {Tuli}}]{Zachrisson2011}
{Zackrisson}, E., {Rydberg}, C.-E., {Schaerer}, D., {{\"O}stlin}, G., \&
  {Tuli}, M. 2011, \apj, 740, 13

\end{thebibliography}
\end{document}